\documentclass[fleqn,10pt]{wlscirep}
\usepackage[utf8]{inputenc}
\usepackage[T1]{fontenc}

\usepackage{amssymb}
\usepackage{amsmath}
\usepackage{lineno}
\usepackage{graphicx}
\usepackage{booktabs}
\usepackage{placeins}
\usepackage{subcaption}
\usepackage{adjustbox}
\usepackage{multirow}
\usepackage{csquotes}
\usepackage{hyperref}

\newcommand{\abbreviations}[1]{{\\ \scriptsize (Abbr.: #1)}}

\newcommand{\SDGradient}{26.2}
\newcommand{\SDGradientSIGN}{13.9}
\newcommand{\SDSignMin}{5.3}
\newcommand{\SDSignMax}{14.1}

\newcommand{\NumSuppTabPerf}{S1}   
\newcommand{\NumSuppTabBoot}{S2}   
\newcommand{\NumSuppTabMask}{S3}   
\newcommand{\NumSuppFigHist}{S1}   
\newcommand{\NumSuppFigLocal}{S2}  

\newcommand{\NumMainFigLocalAime}{1}  
\newcommand{\NumMainFigPatterns}{2}   
\newcommand{\NumMainTabCombined}{1}   

\newcommand{\SuppMethods}{Supplementary Methods}
\newcommand{\SuppTabPerf}{Supplementary Table~\NumSuppTabPerf}
\newcommand{\SuppTabBoot}{Supplementary Table~\NumSuppTabBoot}
\newcommand{\SuppTabMask}{Supplementary Table~\NumSuppTabMask}
\newcommand{\SuppFigHist}{Supplementary Fig.~\NumSuppFigHist}
\newcommand{\SuppFigLocal}{Supplementary Fig.~\NumSuppFigLocal}

\makeatletter
\newcommand{\XRef@check}[3]{%
	\edef\XRef@want{#2}%
	\edef\XRef@got{#3}%
	\ifx\XRef@want\XRef@got\else
		\PackageError{supplementary_refs}%
		{#1 number mismatch: supplementary_refs.tex declares \XRef@want, LaTeX assigned \XRef@got}%
		{A float was added, removed or reordered. Update the numbers in
			supplementary_refs.tex (and recompile both main.tex and
			supplementary.tex) so that the cross-document references stay correct.}%
	\fi}
\newcommand{\CheckSuppFigure}[1]{\XRef@check{Supplementary figure}{#1}{\thefigure}}
\newcommand{\CheckSuppTable}[1]{\XRef@check{Supplementary table}{#1}{\thetable}}
\newcommand{\CheckMainFigure}[1]{\XRef@check{Main-text figure}{#1}{\thefigure}}
\newcommand{\CheckMainTable}[1]{\XRef@check{Main-text table}{#1}{\thetable}}
\makeatother

\title{Beyond Local Inspection: Global, Guideline-Grounded Evaluation of Post-hoc XAI Methods for ECG Classification}

\author[1,2,*]{Nils Gumpfer}
\author[1,2]{Michael Guckert}
\author[3]{Samuel Sossalla}
\author[3]{Birgit A{\ss}mus}
\author[1,2,*]{Jennifer Hannig}

\affil[1]{Hessian Center for Artificial Intelligence (hessian.AI), Darmstadt, Germany}
\affil[2]{Technische Hochschule Mittelhessen, University of Applied Sciences, Friedberg (Hesse), Germany}
\affil[3]{Department of Internal Medicine I, Cardiology, Justus-Liebig-University Giessen, Giessen, Germany}
\affil[*]{nils.gumpfer@kite.thm.de, jennifer.hannig@kite.thm.de}


\begin{abstract}
	Explainable AI (XAI) is used to assess whether artificial intelligence models rely on meaningful patterns, yet explanations that appear plausible for individual predictions may systematically misrepresent model behavior. This is particularly problematic in medicine, where models may rely on irrelevant signal characteristics rather than disease-specific patterns without being recognizable. We address this challenge using electrocardiogram (ECG) data, for which clinical guidelines provide explicit knowledge about diagnostically relevant signal regions. We introduce a global, guideline-grounded framework that aggregates explanations across heartbeats to evaluate them against clinically defined regions of interest. Using four binary classifiers trained on PTB-XL, we assess 13 gradient-based methods across two categories of patterns: low-amplitude segments and high-amplitude QRS morphology. Our results reveal a systematic failure of methods transferred from computer vision. Their explanations often follow signal amplitude rather than clinical relevance, with mean Spearman correlations up to 0.69, leading them to overlook diagnostically decisive low-amplitude regions. For ischemia, LRP-$\epsilon$ assigns only 4.6\% of relevance to the ST segment, compared with 63.8\% for LRP-SIGN. Nine of 13 methods fall below chance for at least one condition, indicating inconsistent reliability across patterns. These findings show that global, domain-grounded evaluation can uncover systematic explanation failures not obvious from sample-level heatmaps.
\end{abstract}

\begin{document}
	
	\flushbottom
	\maketitle
	\thispagestyle{empty}
	
	
	
	\section*{Introduction}
	\FloatBarrier
	
	Artificial intelligence (AI) is increasingly used to support clinical decision-making, with reported performance reaching or even exceeding that of human experts \cite{Rajpurkar2022,Jiang2022}. Nevertheless, adoption in routine practice remains limited, as the opacity of model decisions, together with concerns about reproducibility, robustness, and generalizability, continues to impede trust \cite{Carter2019,Asan2020,Ball2023}. In safety-critical domains such as healthcare, reliable adoption depends not on predictive accuracy alone, but also on model behavior that can be reliably explained, audited, and corrected \cite{Asan2020}, as undesired behaviors acquired during training may remain undetected until and beyond deployment \cite{Lapuschkin2019,Narla2018}. For example, electrocardiogram (ECG) analysis, one of the most widely used, cost-effective, and noninvasive diagnostic tools in cardiology \cite{Guytonhall2011NormECG}, is an active area of clinical AI research \cite{Manimaran2025} with high predictive performance \cite{Attia2019,Hannun2019,Strodthoff2019,GCB2020Paper}. Yet, its practical implementation remains limited, with a lack of trust representing a key barrier despite continued advances in predictive accuracy \cite{Arends2025}.
	
	Because most high-performing medical models are not interpretable by design, trust can develop through repeated evaluation of whether models make use of established diagnostic criteria \cite{Swartout1993,Makovi2023}, for example, through the use explainable AI (XAI). Most commonly, this is done by using local \emph{post-hoc} methods that attribute relevance to input regions driving individual predictions \cite{Samek2021,Holzinger2022,Markus2021} and post-hoc XAI has thus become a primary instrument for auditing model behavior. However, its conclusions depend strongly on the chosen attribution method, whose reliability remains contested \cite{Payrovnaziri2020}. Owing to their differing computational principles, attribution methods can produce substantially different explanations for the same prediction \cite{Ali2023}. Moreover, an explanation that fails to capture the features actually driving a prediction may be misleading while still appearing plausible. This risk is compounded by the common practice of assessing explanation validity through visual inspection of individual examples, where plausible-looking explanations are often interpreted as evidence of clinically meaningful model behavior \cite{Strodthoff2019,Goettling2024}. Yet, visual plausibility primarily reflects agreement with human expectations rather than faithfulness to the model's internal decision process \cite{xaimanifesto,Guckert2021}.
	
\begin{figure}[t]
		\centering
		\noindent
		\includegraphics[width=\linewidth]{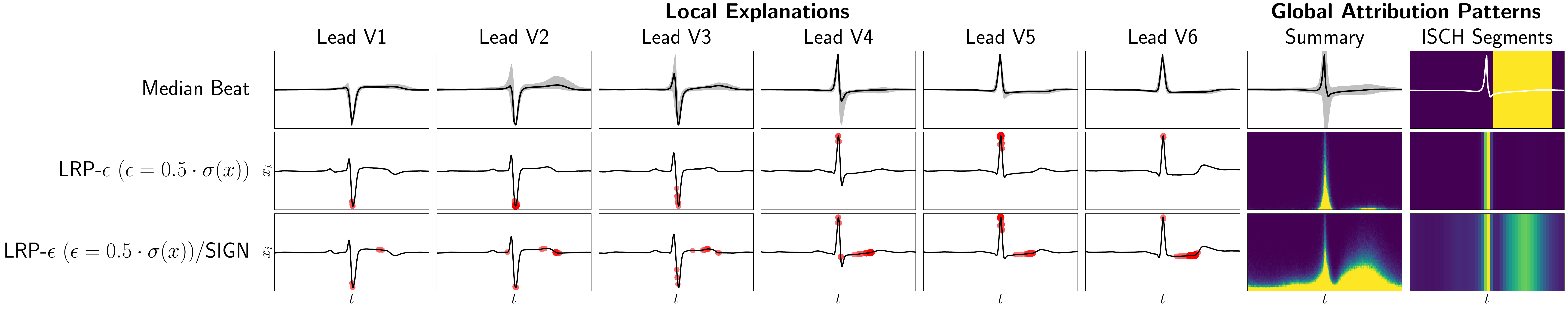}
		\caption{Local vs.\ global attribution patterns for myocardial ischemia, shown for LRP-$\epsilon$ \cite{Bach2015} and its SIGN-adjusted \cite{Gumpfer2023} variant. The left side shows the local explanation of a single R-peak-centered heartbeat from an individual recording \cite{AIME2024Paper} in the precordial leads V1-V6, where positive relevance $R_i^+$ is overlaid as red dots whose size and color encode the attributed relevance. The top row displays the median beat with its interquartile range (grey) alongside the guideline-based reference mask \cite{AHA2009b} (Fig.~\ref{fig:patterns_criteria}). The right side shows the global attribution patterns, namely the relevance histograms summarized across all leads and their segment-wise aggregations, compared against the ISCH reference mask. The plotted example shows that the local attribution patterns (especially in leads V4-V6) for both methods are no exception but rather are in line with the global attribution patterns present over multiple recordings, a correspondence that only the aggregated view can establish. The same overview for all four examined pathologies is provided as \SuppFigLocal. \abbreviations{ISCH = Myocardial ischemia}}
		\label{fig:localaime}
		\CheckMainFigure{\NumMainFigLocalAime}
	\end{figure}

	Evaluating explanation methods is therefore an open problem in its own right. No universal metric for explanation quality exists, and there is often no defined ground truth for what a model should attend to, so most studies either omit thorough evaluation or rely on subjective judgment of isolated cases \cite{Payrovnaziri2020,Salih2024}, leaving open how well explanations meet the requirements of clinicians \cite{Markus2021,Langlais2023,Elkhawaga2024}. A single prediction moreover cannot reveal whether an attribution pattern reflects systematic behavior across a dataset, a model, or an explanation method, or merely an instance-specific effect \cite{AIME2024Paper}. Figure~\ref{fig:localaime} illustrates this for two methods explaining the same ischemia prediction, where the aggregated view shows how each distributes relevance relative to the diagnostic ST segment across recordings, and \SuppFigLocal\ extends this to all four examined pathologies.

	ECG analysis is well-suited to examine this problem, as its clinical interpretation relies on well-established temporal and morphological criteria, such as PR-interval prolongation, ST-segment deviation, and characteristic QRS shapes, codified in international guidelines and applied routinely in practice \cite{AHA2008,AHA2009a,AHA2009b}. For each condition, these guidelines pinpoint the signal regions a clinician inspects, and therefore the regions a trustworthy explanation should emphasize, yielding a domain-specific ground truth rarely available in other settings. Because diagnosis rests on recurring morphological patterns, this ground truth is naturally expressed at the dataset level and best exploited by aggregating explanations across many samples, which reveals general attribution behavior hidden in local views \cite{Samek2019,Lapuschkin2019}.

	A further motivation is a concern widely suspected but hard to confirm locally, namely that gradient-based methods have repeatedly been reported to emphasize high-amplitude ECG components across diagnostic tasks \cite{Goettling2024,Metsch2025,Strodthoff2019,Najjar2023,Guckert2021,Wagner2024,Strodthoff2021,AIME2024Paper}. Because ECG amplitudes differ markedly between waveform components, most notably the dominant R-peak, these methods may over-attribute large-amplitude regions even when the diagnostically critical information lies in low-amplitude regions such as the PR interval or the ST segment \cite{Najjar2023,Goettling2024}. Whether this reflects a systematic tendency or only isolated cases cannot be derived from the local samples and small case studies so far \cite{AIME2024Paper}, but requires dataset-level evaluation against guideline-defined expectations, as in the aggregated view of Fig.~\ref{fig:localaime} (right part).
	
	Prior work approaches this gap without closing it. Wagner et al.~\cite{Wagner2024} aggregated local attribution maps across patient subgroups on PTB-XL~\cite{Wagner2020a}, yet compared methods only indirectly through amplitude regression tasks, and earlier studies \cite{VanDeLeur2021,Bender2024} relied on beat-aligned visualization assessed by purely qualitative inspection, so that attributions are summarized or displayed but never measured against an external clinical reference. A similar gap exists on the method side, as the SIGN approach \cite{Gumpfer2023}, introduced on images to correct the contrast bias arising when gradients are multiplied by their inputs \cite{Ancona2018}, has been examined on physiological time series, where amplitude carries a different meaning, only for individual local examples \cite{AIME2024Paper}. Recent reviews \cite{Salih2024,Manimaran2025,Gumpfer2026} likewise note that ECG and time-series XAI is often restricted to a single local method, rarely standardized in its evaluation, and frequently borrowed from computer vision without adaptation. What remains missing is a quantitative, guideline-grounded evaluation comparing attribution patterns directly against expert-defined diagnostic regions across a broad range of gradient-based methods.

	We therefore present a global, guideline-grounded framework for assessing the clinical plausibility of explanations in ECG classification. Rather than judging single predictions, it aggregates explanations across many heartbeats and compares them against guideline-derived reference regions, turning established diagnostic criteria into an objective benchmark for explanation quality. It applies in two complementary ways, as it can compare explanation methods to identify which produce clinically plausible attributions, but also models, architectures, or training conditions to audit what a given model has learned. Throughout our experiments, the model serves as a fixed pattern proxy, so that we assess plausibility against the guideline rather than faithfulness to the model. Our main contributions are summarized below:
	
	\begin{itemize}
		\item A reusable, open-source evaluation framework (see \textit{Data Availability}) that combines aggregated relevance distributions, amplitude-relevance correlation analysis, and expert-defined reference masks into a domain-specific ground truth, addressing the lack of standardized, human-centered benchmarks in time-series XAI. Because clinical use demands reliability on every condition, the framework scores methods by worst-case rather than average agreement with the guideline.
		\item A systematic comparison of 13 gradient-based attribution methods on CNNs trained on the PTB-XL dataset \cite{Wagner2020a} across four cardiac pathologies using guideline-derived reference masks \cite{AHA2008,AHA2009a,AHA2009b}. These cover two categories of ECG diagnostic criteria, low-amplitude intervals / segments near the isoelectric line (atrioventricular block, myocardial ischemia) and high-amplitude QRS morphology patterns (right and left bundle branch block). We show that methods transferred directly from computer vision are strongly driven by signal amplitude and systematically overlook diagnostically important regions such as the PR and ST segments.
		\item Evidence that lightweight, domain-aware adaptations like SIGN \cite{Gumpfer2023} for time-series data restore consistent coverage of guideline-defined diagnostic regions across all examined pathologies at low computational overhead.
	\end{itemize}
	
	\FloatBarrier
	\section*{Results} 
	\FloatBarrier
	
	We systematically evaluated 13 gradient-based post-hoc explanation methods applied to four independently trained binary CNNs, one per pathology, covering four clinically distinct cardiac conditions, namely atrioventricular block (AVB), myocardial ischemia (ISCH), right bundle branch block (RBBB), and left bundle branch block (LBBB). These span two categories of ECG diagnostic criteria, low-amplitude intervals and segments near the isoelectric line (PR interval for AVB, ST segment for ISCH) and high-amplitude QRS morphology patterns (RBBB, LBBB), therefore covering both amplitude ranges in which attribution behavior may diverge. All four models achieved high classification performance on the PTB-XL~\cite{Wagner2020a} benchmark dataset (see \SuppTabPerf), ensuring that subsequent analyses assessed explanation quality rather than model accuracy limitations.
	
	\subsection*{Global attribution patterns}
	
	The aggregated relevance histograms (Fig.~\ref{fig:hist_summary} left and \SuppFigHist) revealed that several methods consistently concentrated relevance on high-amplitude waveform components, particularly the R-peak and T-wave, independently of pathology-specific diagnostic regions, so that their patterns differed strongly from the guideline-derived reference masks (Fig.~\ref{fig:patterns_criteria}), leaving the masked segments largely unattributed. Methods involving input multiplication or baseline-dependent contrast shifts, such as Gradient~$\times$~Input~\cite{Shrikumar2017}, SmoothGrad~$\times$~Input~\cite{Smilkov2017}, Integrated Gradients~\cite{Sundararajan2017}, DeepSHAP~\cite{Lundberg2017}, GradSHAP~\cite{Erion2021}, and LRP-$\epsilon$~\cite{Bach2015}, aligned visibly more closely with the input baseline, their histograms following the amplitude shape of the median beat across all four pathologies rather than the respective diagnostic segments. The remaining methods, namely VarGrad, Random, SmoothGrad, plain Gradient, and the SIGN-adjusted variants, distributed relevance more evenly across the cardiac cycle, with histograms shifting between pathologies toward the respective reference regions rather than repeating the same amplitude-tracking shape throughout, a distinction detailed per pathology further below.

	\subsection*{Correlation between relevance and input}
	
	These observations were confirmed quantitatively by a correlation analysis (Table~\ref{tab:combined_analysis} and Fig.~\ref{fig:hist_summary} right) using Spearman's Rank Correlation Coefficient (SCC). The input baseline produced the expected $\overline{\text{SCC}} = 1.00$, while the highest mean SCC values among XAI methods were observed for Integrated Gradients (0.69), LRP-$\epsilon$ ($\epsilon = 0.5 \cdot \sigma(x)$) (0.67), SmoothGrad~$\times$~Input (0.65), and Gradient~$\times$~Input (0.57), all of which correlated strongly and positively with absolute input in every individual pathology (SCC between 0.24 and 0.76), indicating relevance predominantly driven by signal amplitude rather than model behavior.
	
	In contrast, VarGrad~\cite{Adebayo2018}, Random, SmoothGrad~\cite{Smilkov2017}, SmoothGrad~$\times$~SIGN~\cite{Gumpfer2023}, Gradient~\cite{Simonyan2014}, Gradient~$\times$~SIGN~\cite{Gumpfer2023}, and LRP-$\epsilon$/SIGN~\cite{Gumpfer2023} exhibited markedly lower correlations ($\overline{\text{SCC}}$ between $-0.24$ and $0.16$). SIGN-based variants in particular remained below an SCC of 0.4 in every individual pathology, with per-pathology maxima of 0.39 for Gradient~$\times$~SIGN, 0.39 for LRP-$\epsilon$/SIGN, and 0.15 for SmoothGrad~$\times$~SIGN. Their correlations also changed sign across pathologies (e.g., $-0.34$ for Gradient~$\times$~SIGN on AVB vs. 0.39 on LBBB), whereas the input-multiplied methods stayed positive throughout.
	
	\begin{table*}
		\centering
		\caption{Combined analysis of correlation and pattern coverage across explanation methods and pathologies. Spearman's rank correlation coefficient (SCC; equation~\eqref{eq:scc}) was computed between absolute relevance and absolute input amplitudes. Coverage was computed using the guideline-based reference masks from Fig.~\ref{fig:patterns_criteria} and equation~\eqref{eq:cov}. Normalized coverage ($NCov$; equation~\eqref{eq:ncov}) expresses coverage relative to the random baseline as a skill score, where $0\,\%$ denotes chance level, $100\,\%$ perfect attribution, and negative values worse-than-chance attribution. Worst-case normalized coverage ($NCov_{min}$; equation~\eqref{eq:cons}) is the minimum NCov across all examined pathologies. Rows are sorted by $NCov_{min}$ in descending order, so that methods remaining most reliable on their most challenging pathology appear at the top, and the random baseline ($NCov_{min} = 0\,\%$) separates methods that stay above chance on every pathology from those that fall below chance on at least one. Within each column, the highest value is marked in bold and the lowest in italics; the two baselines (Input, Random) are included in this comparison.
			\abbreviations{AVB = AV-Block, GT = Ground truth, ISCH = Myocardial ischemia, RBBB = Right bundle branch block, LBBB = Left bundle branch block, SCC = Spearman's rank correlation coefficient, NCov = Normalized coverage, $NCov_{min}$ = Worst-case NCov}}
		\label{tab:combined_analysis}
		\CheckMainTable{\NumMainTabCombined}
		
		\resizebox{\textwidth}{!}{%
			\begin{tabular}{lcccccccccccccccc}
\toprule
& \multicolumn{5}{c}{\textbf{Correlation Analysis (SCC)}} & \multicolumn{11}{c}{\textbf{Pattern Coverage Analysis}} \\
\cmidrule(lr){2-6}
\cmidrule(lr){7-17}
\textbf{Method} & \multicolumn{4}{c}{\textbf{Pathology}} & \textbf{Mean} & \multicolumn{4}{c}{\textbf{Coverage (GT, \%)}} & \textbf{Mean} & \multicolumn{4}{c}{\textbf{Normalized (NCov, \%)}} & \textbf{Mean} & \textbf{$NCov_{min}$} \\
\cmidrule(lr){2-5}
\cmidrule(lr){7-10}
\cmidrule(lr){12-15}
& \textbf{AVB} & \textbf{ISCH} & \textbf{RBBB} & \textbf{LBBB} & & \textbf{AVB} & \textbf{ISCH} & \textbf{RBBB} & \textbf{LBBB} & & \textbf{AVB} & \textbf{ISCH} & \textbf{RBBB} & \textbf{LBBB} & & \\
\midrule
LRP-$\epsilon~(\epsilon = 0.5 \cdot \sigma(x))$/SIGN & -0.21 & ~~0.24 & ~~0.21 & ~~0.39 & ~~0.16 & 71.4 & \textbf{63.8} & 87.0 & 75.3 & \textbf{74.4} & 56.7 & \textbf{41.5} & 80.9 & 63.7 & \textbf{60.7} & \textbf{41.5} \\
SmoothGrad $\times$ SIGN & -0.04 & ~~0.15 & ~~0.03 & -0.01 & ~~0.03 & 58.9 & 52.2 & 51.9 & 51.0 & 53.5 & 37.7 & 22.9 & 29.4 & 28.0 & 29.5 & 22.9 \\
Gradient $\times$ SIGN & \textit{-0.34} & -0.09 & ~~0.34 & ~~0.39 & ~~0.08 & 65.5 & 52.2 & 73.1 & 66.3 & 64.3 & 47.7 & 22.8 & 60.4 & 50.5 & 45.4 & 22.8 \\
Gradient $\times$ Input & ~~0.24 & ~~0.65 & ~~0.67 & ~~0.71 & ~~0.57 & 56.9 & 47.7 & 90.0 & 85.7 & 70.1 & 34.7 & 15.5 & 85.4 & 79.0 & 53.6 & 15.5 \\
Random & ~~0.00 & ~~0.00 & ~~0.01 & ~~0.00 & ~~0.00 & 34.0 & 38.1 & 31.9 & \textit{31.9} & \textit{34.0} & ~~0.0 & ~~0.0 & ~~0.0 & ~~\textit{0.0} & ~~\textit{0.0} & ~~0.0 \\
SmoothGrad & -0.05 & ~~0.13 & ~~0.02 & -0.02 & ~~0.02 & 54.5 & 36.3 & 40.3 & 50.8 & 45.5 & 31.0 & -2.9 & 12.3 & 27.8 & 17.1 & -2.9 \\
GradSHAP & ~~0.25 & ~~0.38 & ~~0.30 & ~~0.37 & ~~0.33 & 55.5 & 34.7 & 87.0 & 83.5 & 65.2 & 32.6 & -5.5 & 80.9 & 75.8 & 45.9 & -5.5 \\
LRP-$\alpha\beta$ & ~~0.12 & ~~0.58 & ~~0.37 & ~~0.66 & ~~0.43 & 74.1 & 33.8 & 62.2 & \textbf{95.5} & 66.4 & 60.8 & -6.9 & 44.5 & \textbf{93.3} & 47.9 & -6.9 \\
Gradient & \textit{-0.34} & -0.09 & ~~0.34 & ~~0.38 & ~~0.07 & 64.0 & 32.8 & 68.1 & 69.6 & 58.6 & 45.4 & -8.6 & 53.1 & 55.3 & 36.3 & -8.6 \\
VarGrad & -0.10 & \textit{-0.16} & \textit{-0.28} & \textit{-0.41} & \textit{-0.24} & \textbf{77.4} & 38.6 & \textit{24.3} & 36.0 & 44.1 & \textbf{65.8} & ~~0.9 & \textit{-11.2} & ~~6.0 & 15.4 & -11.2 \\
Integrated Gradients & ~~0.62 & ~~0.72 & ~~0.68 & ~~0.73 & ~~0.69 & 28.1 & 29.0 & 96.2 & 80.6 & 58.5 & -9.0 & -14.6 & 94.4 & 71.5 & 35.6 & -14.6 \\
Input & ~~\textbf{1.00} & ~~\textbf{1.00} & ~~\textbf{1.00} & ~~\textbf{1.00} & ~~\textbf{1.00} & \textit{12.2} & 35.1 & 60.7 & 46.5 & 38.6 & \textit{-33.1} & -4.8 & 42.3 & 21.5 & ~~6.5 & -33.1 \\
SmoothGrad $\times$ Input & ~~0.40 & ~~0.75 & ~~0.74 & ~~0.71 & ~~0.65 & 29.0 & 17.4 & 87.5 & 87.6 & 55.4 & -7.6 & -33.4 & 81.6 & 81.8 & 30.6 & -33.4 \\
DeepSHAP & ~~0.20 & ~~0.31 & ~~0.36 & ~~0.42 & ~~0.32 & 58.3 & 14.7 & 91.1 & 93.6 & 64.4 & 36.9 & -37.7 & 86.9 & 90.6 & 44.2 & -37.7 \\
LRP-$\epsilon~(\epsilon = 0.5 \cdot \sigma(x))$ & ~~0.47 & ~~0.73 & ~~0.73 & ~~0.76 & ~~0.67 & 50.4 & ~~\textit{4.6} & \textbf{97.0} & 93.9 & 61.5 & 24.9 & \textit{-54.0} & \textbf{95.6} & 91.0 & 39.4 & \textit{-54.0} \\
\bottomrule
\end{tabular}%
		}
	\end{table*}

	\subsection*{Alignment with clinically relevant patterns}
	
\begin{figure}[p]
		\centering
		\begin{subfigure}[t]{\textwidth}
			\centering
			\includegraphics[height=6cm]{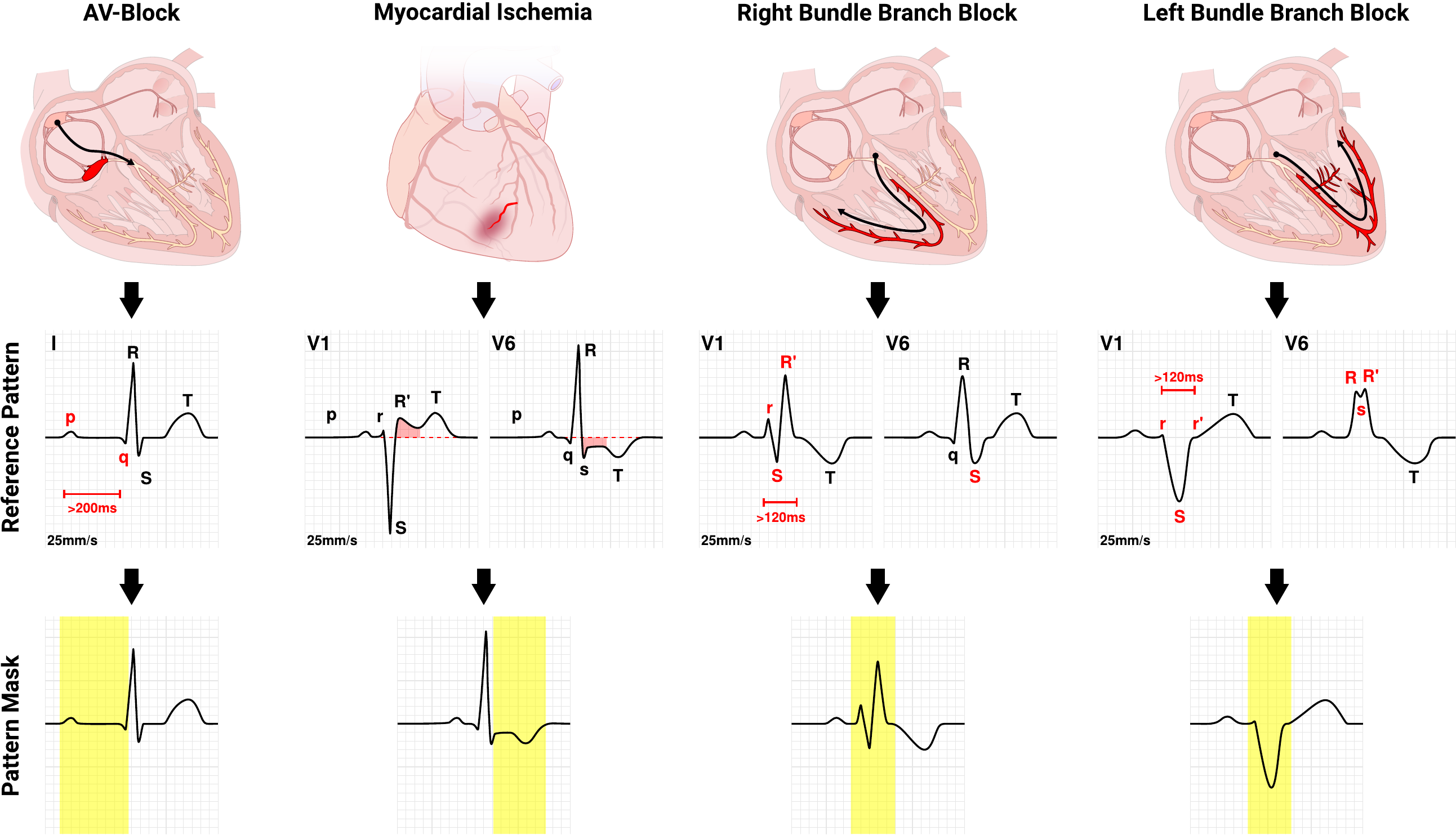}\\[2mm]
			\caption{Guideline-informed reference masks for the four pathologies.}
			\label{fig:patterns_criteria}
		\end{subfigure}
		
		\vspace{1em}
		
		\begin{subfigure}[t]{\textwidth}
			\centering
			\noindent
			\adjustbox{valign=t}{\includegraphics[height=13.2cm,trim=0 0 4750 0,clip]{hist_AVBLOCK.pdf}}\hspace{1mm}%
			\adjustbox{valign=t}{\includegraphics[height=13.2cm,trim=5061 0 0 0,clip]{hist_AVBLOCK.pdf}}\hspace{1mm}%
			\adjustbox{valign=t}{\includegraphics[height=13.2cm,trim=5061 0 0 0,clip]{hist_ISCH.pdf}}\hspace{1mm}%
			\adjustbox{valign=t}{\includegraphics[height=13.2cm,trim=5061 0 0 0,clip]{hist_RBBB.pdf}}\hspace{1mm}%
			\adjustbox{valign=t}{\includegraphics[height=13.2cm,trim=5061 0 0 0,clip]{hist_LBBB.pdf}}\hspace{1mm}%
			\adjustbox{valign=t}{\includegraphics[height=13.2cm,trim=620 0 0 0,clip]{correlation.png}}%
			\caption{Aggregated relevance histograms (left) and amplitude-relevance correlation plots (right), per method and pathology.}
			\label{fig:hist_summary}
		\end{subfigure}

		\caption{Guideline-informed reference masks and the global attribution patterns evaluated against them. Panel \textbf{a} marks the regions that diagnostic guidelines \cite{AHA2008,AHA2009a,AHA2009b} designate for each pathology, namely PR-interval prolongation for AV-block, ST-segment deviation for ischemia, and QRS morphology in V1-V2 and V5-V6 for the bundle branch blocks. The masks encode only the temporal extent of these criteria and apply across all leads (see Reference Patterns). Panel \textbf{b} compares each method's aggregated relevance histograms against those masks, quantified as coverage in Table~\ref{tab:combined_analysis} and equation~\eqref{eq:cov}, and correlates absolute relevance with absolute input amplitude, showing how far a method follows amplitude rather than the diagnostic region. \abbreviations{AVB = AV-Block, ISCH = Myocardial ischemia, RBBB = Right bundle branch block, LBBB = Left bundle branch block}}
		\label{fig:masks_and_patterns}
		\CheckMainFigure{\NumMainFigPatterns}
	\end{figure}

	The correlation analysis quantifies amplitude dependence, but not whether attributions land in the regions clinicians consult for diagnosis. The pathology-specific reference masks \cite{AHA2008,AHA2009a,AHA2009b} (Fig.~\ref{fig:patterns_criteria}) provide this criterion, and the aggregated relevance histograms (Fig.~\ref{fig:hist_summary} left and \SuppFigHist) allow alignment to be assessed for every method, beat position, and lead, revealing pathology-specific deviations that global correlation values cannot resolve.
	
	\paragraph*{Atrioventricular block.} The diagnostically relevant region is the PR interval (Fig.~\ref{fig:patterns_criteria}, leftmost panel), which extends from P-wave onset to QRS onset and, apart from the low-amplitude P-wave, remains close to the isoelectric line. The AVB histograms reveal that the input-multiplied methods concentrate relevance almost exclusively on the R-peak and, to a lesser extent, the T-wave, leaving the PR interval essentially unattributed, and their aggregated patterns closely resemble the input baseline. SIGN-adjusted methods and, to a lesser extent, plain Gradient distribute relevance more broadly across the cardiac cycle and assign visibly higher relevance to the PR interval.
	
	\paragraph*{Myocardial ischemia.} The diagnostically relevant region is the ST segment in precordial leads (V1-V6, Fig.~\ref{fig:patterns_criteria}, second panel), again a low-amplitude segment lying between two high-amplitude landmarks (R-peak and T-wave). The ISCH histograms show that for amplitude-sensitive methods relevance accumulates strongly on R-peaks and T-waves in V1-V6, while the intervening ST segment receives little to no attribution. The effect is most pronounced for LRP-$\epsilon$ ($\epsilon = 0.5 \cdot \sigma(x)$), SmoothGrad~$\times$~Input, and Integrated Gradients, whose precordial histograms are visually almost indistinguishable from the input baseline. Plain Gradient, despite its low amplitude correlation, does not reliably highlight the ST segment either, its histograms in V1-V6 showing relevance scattered broadly along the cardiac cycle. SIGN-adjusted methods, in contrast, retain measurable relevance on the ST segment across V1-V6.
	
	\paragraph*{Right and left bundle branch block.} Both bundle branch blocks are characterized by widened QRS complexes with distinctive morphologies, an \enquote{M}-shape in V1-V2 for RBBB and a deep S followed by a broad, notched R in V5-V6 for LBBB (Fig.~\ref{fig:patterns_criteria}, third and fourth panels). In contrast to AVB and ISCH, their diagnostic regions coincide with high-amplitude waveform components. The RBBB and LBBB histograms show that most methods concentrate substantial relevance on the QRS complex in the relevant lead pairs, but that amplitude-sensitive methods collapse it onto a single point, typically the R-peak, rather than spreading it across the morphological pattern that distinguishes them from a normal QRS complex. SIGN-adjusted methods distribute relevance more evenly along the QRS complex, capturing the shape rather than only its highest-amplitude point. This distinction does not always translate into a coverage difference, since the mask covers the entire QRS region, and is visible only in the spectral structure of the histograms.
	
	\paragraph*{Cross-pathology pattern.} The cross-pathology summary in Fig.~\ref{fig:hist_summary} shows that SIGN-adjusted methods produce histograms varying clearly between pathologies and shifting toward the respective reference regions, whereas the input-multiplied histograms stay visibly similar throughout, although the reference masks differ substantially. Differences that appear modest in summary statistics are thus clearly resolved in the spectral histograms.
	
	\subsection*{Coverage and worst-case reliability across pathologies}
	
	A coverage analysis (Table~\ref{tab:combined_analysis}) quantified relevance assigned to guideline-defined diagnostic regions, with interpretation focusing on normalized coverage (NCov), which expresses attribution agreement relative to the random baseline (NCov~=~0\,\%). The input baseline achieved only NCov~=~6.5\,\%, confirming that ECG amplitude alone provides little diagnostic agreement.
	
	The results revealed a clear distinction between high average performance and stable behavior across pathologies. The same input-multiplied and baseline-dependent methods achieved high coverage where diagnostic regions overlap with high-amplitude waveform components, particularly RBBB and LBBB, but performed poorly on myocardial ischemia, where the diagnostically relevant ST segment has comparatively low amplitude. LRP-$\epsilon$ with $\epsilon = 0.5 \cdot \sigma(x)$, for example, reached NCov~=~95.6\,\% on RBBB and 91.0\,\% on LBBB, yet dropped to $-$54.0\,\% on ISCH, and SmoothGrad~$\times$~Input, Integrated Gradients, and DeepSHAP likewise turned negative on ISCH despite comparatively high average coverage. Their elevated mean performance therefore reflects favorable alignment with high-amplitude diagnostic regions rather than robust agreement with clinical reasoning across pathologies, and the same holds to a lesser extent for AVB, whose low-amplitude PR interval poses a similar challenge.
	
	This fragility is directly captured by the worst-case normalized coverage ($NCov_{min}$; equation~\eqref{eq:cons}). Methods dominated by input amplitude dependence showed strongly negative $NCov_{min}$ values, with LRP-$\epsilon$ ($\epsilon = 0.5 \cdot \sigma(x)$) reaching $-54.0$\,\% and several others below $-5\,\%$. Even plain Gradient, which exhibited low amplitude correlation, dropped below chance on ISCH (NCov~=~$-8.6\,\%$), confirming that reduced amplitude dependence alone does not guarantee reliable attribution.

	In contrast, four methods stayed above chance on every pathology, namely the three SIGN-based variants and Gradient~$\times$~Input (15.5\,\%), the SIGN family being the only one whose members were all positive. LRP-$\epsilon$/SIGN with $\epsilon = 0.5 \cdot \sigma(x)$ reached the highest worst-case coverage by a clear margin ($NCov_{min}$~=~41.5\,\%, 95\,\% CI [40.0, 43.1]), followed by SmoothGrad~$\times$~SIGN (22.9\,\%, [21.6, 24.2]) and Gradient~$\times$~SIGN (22.8\,\%, [20.5, 25.0]), the latter two statistically indistinguishable, as their confidence intervals overlap almost entirely (\SuppTabBoot). The worst case for all three SIGN methods was ISCH, yet their $NCov_{min}$ values remained well above zero, indicating that the SIGN modification preserves attribution in low-amplitude diagnostic regions under the most challenging conditions. LRP-$\epsilon$/SIGN with $\epsilon = 0.5 \cdot \sigma(x)$ also achieved the highest mean normalized coverage overall ($\overline{NCov}$~=~60.7\,\%), ahead of Gradient~$\times$~Input (53.6\,\%), while the remaining SIGN variants reached 45.4\,\% (Gradient~$\times$~SIGN) and 29.5\,\% (SmoothGrad~$\times$~SIGN).

	The pathology determining $NCov_{min}$ was highly consistent, falling on ISCH for 12 of the 13 methods, the only exception being VarGrad, whose minimum fell on RBBB ($-11.2\,\%$). For the input baseline, the minimum occurred on AVB ($-33.1\,\%$). Mean and worst-case coverage therefore ranked methods differently, as DeepSHAP, for example, ranked sixth by $\overline{NCov}$ (44.2\,\%) but second to last by $NCov_{min}$ ($-37.7\,\%$).

	To assess robustness, we conducted two supplementary analyses. First, 95\,\% bootstrap confidence intervals (1000 resamples at the recording level) confirmed that the $NCov_{min}$ ordering among method groups is statistically stable, the intervals for all three SIGN-based methods lying entirely above zero while those for the amplitude-sensitive methods remain largely negative (\SuppTabBoot). Second, a mask sensitivity analysis in which the mask boundaries were shifted symmetrically by $k \in \{-20, -10, 0, +10, +20\}$~timesteps showed that the qualitative ranking is preserved at moderate offsets ($|k| \leq 10$~timesteps), where SIGN-based methods consistently achieve positive $NCov_{min}$ while most input-multiplied methods remain negative on at least one pathology. At the largest dilation ($k = +20$~timesteps), several amplitude-sensitive methods flip to positive $NCov_{min}$ because the expanded mask engulfs the high-amplitude QRS region, removing the basis of the distinction, yet LRP-$\epsilon$/SIGN retains the highest $NCov_{min}$ at every offset (\SuppTabMask).

	\FloatBarrier
	\section*{Discussion}

	Applied to the same four models and the same true-positive predictions, the 13 explanation methods produced markedly different relevance maps. Methods that multiply the gradient by the input, or that subtract a baseline, concentrated relevance on the R-peak and T-wave irrespective of the pathology and followed signal amplitude closely ($\overline{\text{SCC}}$ up to 0.69), while the SIGN-adjusted variants stayed near zero (0.03-0.16) and shifted their attribution toward the guideline region of each condition. The input-multiplied methods therefore track the signal envelope whereas the SIGN variants track the pathology, a difference that becomes visible only through global, guideline-grounded evaluation, since a single explanation cannot show whether a pattern is systematic.

	Averaged over pathologies, this behavior is easily missed, since several methods achieved strong overlap for bundle branch blocks, where diagnostic regions coincide with high-amplitude waveforms, yet performed considerably worse on myocardial ischemia, where the low-amplitude ST segment reached only 4.6\,\% coverage for LRP-$\epsilon$ ($\epsilon = 0.5 \cdot \sigma(x)$). The worst-case normalized coverage captures this fragility and matches deployment conditions, where a method is chosen once and then applied to whichever condition arises, so that an average would conceal the failing pathology. That the worst case falls on ISCH for almost every method is not an artifact of the criterion but its motivation, since the low-amplitude category is where amplitude-driven attribution breaks down, and which category a clinical application falls into is not known in advance. Alignment with the guideline on three of four pathologies therefore provides no basis for trusting the fourth, and the bundle branch blocks, on which amplitude-sensitive methods appear capable, cannot serve as a benchmark on their own.
	
	\subsection*{Origins of amplitude dependence}
	The observed amplitude-dependent relevance patterns arise from the mathematical structure of the methods (see \SuppMethods). For input-multiplied methods such as Gradient~$\times$~Input, the high SCC with the raw signal is partly true by construction, since multiplying gradients by input values forces $|R_i|$ to co-vary with $|x_i|$ regardless of the model's reasoning, and Integrated Gradients, DeepSHAP, GradSHAP, and LRP-$z$/$\epsilon$ similarly reinforce high-amplitude regions through baseline-dependent contrast shifts. The coverage divergence, where these methods fail on ischemia despite reasonable mean values, is the more diagnostic evidence, since it shows the bias makes attribution miss the low-amplitude regions that define the pathology. This mirrors the contrast effects documented for image data \cite{Gumpfer2023,Ancona2018}, showing that the same bias emerges in 1D physiological time series.

	Within this group, Gradient~$\times$~Input stands out, as despite its strong amplitude correlation ($\overline{\text{SCC}} = 0.57$) it is the only amplitude-sensitive method to maintain positive worst-case normalized coverage ($NCov_{min}$~=~15.5\,\%), whereas Integrated Gradients, DeepSHAP, and LRP-$\epsilon$ with $\epsilon = 0.5 \cdot \sigma(x)$ showed strongly negative values ($-14.6$, $-37.7$, and $-54.0$\,\%), indicating systematic avoidance of the diagnostically relevant region on at least one pathology. Its better floor likely reflects the absence of baseline subtraction, since Integrated Gradients and SHAP-based methods can produce large negative deviations when baseline and input differ markedly. Input multiplication alone is thus less harmful than baseline-dependent contrast amplification.

	The SIGN variants suppress this coupling by replacing the input factor with its sign, discarding magnitude while retaining the directional reference, and their correlations accordingly stayed below 0.4 in every pathology, mirroring the correction originally reported for image data \cite{Gumpfer2023}. Comparing plain Gradient with Gradient~$\times$~SIGN isolates the contribution of the sign reference from that of amplitude debiasing alone. Both exhibit low amplitude correlation ($\overline{\text{SCC}} = 0.07$ and $0.08$), yet differ substantially in cross-pathology stability (SD of NCov = \SDGradient\,\% vs. \SDGradientSIGN\,\%), and on ischemia plain Gradient scatters relevance along the cardiac cycle without concentrating on the ST segment, while the SIGN variants retain relevance there. Reduced amplitude coupling is therefore necessary but not sufficient for clinically aligned attribution. The sign adjustment additionally prevents pathology-dependent cancellation when aggregating across beats with varying morphology, allowing low-amplitude regions to be highlighted even when high-amplitude features lie in immediate temporal proximity. High amplitude-relevance correlation should therefore be read as evidence of insufficient domain adaptation, not of faithfulness to the model.

	\subsection*{Clinical plausibility and model faithfulness}

	Coverage quantifies agreement between an attribution and a guideline-defined region, but does not establish that a method reflects the internal computation of the model, and we did not test that linkage directly. The models instead serve as a pattern proxy, in that they are predictively valid (AUROC 0.93-0.99, \SuppTabPerf), the analysis is confined to true positives, and the same architecture and pathologies were previously shown at lead level to concentrate relevance on the leads that guidelines prescribe \cite{AIME2024Paper}. The comparison itself does not rest on this assumption, since all 13 methods are applied to the identical model per pathology, so that coverage differences are attributable to the methods. On the same recordings, LRP-$\epsilon$/SIGN places 63.8\,\% of its relevance in the ST segment while LRP-$\epsilon$ places 4.6\,\%, and one model cannot account for both.

	A high coverage value nevertheless does not certify faithfulness, since a method could attribute to regions the model does not use, and the SIGN variants show only that guideline-aligned attribution is attainable on these models, which the amplitude-driven methods fail to achieve. The framework likewise cannot arbitrate the case in which all methods agree on a region outside the guideline, where a model shortcut and a shared method bias would be indistinguishable. Guideline coverage is therefore complementary to faithfulness evaluation, not a substitute.

	\subsection*{Implications for practice}

	Clinically meaningful interpretation of ECG signals depends on temporal relationships and segment-specific patterns \cite{Guytonhall2011NormECG,AHA2008,AHA2009a,AHA2009b}. Our results show that several widely used explanation methods do not reliably highlight these features under global evaluation, instead concentrating relevance on high-amplitude peaks that may matter for conditions such as left ventricular hypertrophy but fail to capture the diagnostic patterns of the four pathologies studied here. Critically, this unreliability is not apparent from local visual inspection, the standard evaluation practice in clinical AI research \cite{Markus2021}, so visually convincing local explanations can coexist with systematic under-attribution of diagnostically decisive regions and create false confidence in explanation validity. Methods with reduced amplitude dependence, such as SIGN-based variants, demonstrated more reliable and pathology-consistent attribution across all examined conditions and are a more suitable choice for clinical application.

	\subsection*{Comparison with prior work}
    
	Our results align with prior observations that gradient-based methods overemphasize high-amplitude or visually salient regions \cite{Najjar2023,Strodthoff2019,Guckert2021,Wagner2024,Goettling2024,AIME2024Paper}, and extend the recent dataset-wide analysis of Wagner et al.~\cite{Wagner2024}, who identified plain Gradient as the most reliable of four methods on amplitude correlation. Broadening the comparison to 13 methods and guideline-derived masks, however, reveals that plain Gradient exhibits substantial cross-pathology variability (SD of NCov = \SDGradient\,\%) that is markedly reduced by SIGN-based adjustments (SD of NCov = \SDSignMin-\SDSignMax\,\%), motivating their inclusion in future ECG XAI evaluations. The consistency of amplitude-driven under-attribution across domains supports the view that these limitations reflect a general property of methods transferred from computer vision without adaptation \cite{Gumpfer2023}, and recent work has further shown that some XAI approaches fail to highlight clinically meaningful features \cite{Metsch2025}.
	
	\subsection*{Methodological contribution}

	The proposed framework addresses documented gaps in XAI evaluation for time-series classification \cite{Gumpfer2026}, as it is reusable across methods and models and requires no per-sample attribution ground truth, while its expert-validated masks address both the scarcity of annotated XAI benchmarks and the lack of human-centered validation.
	
	\subsection*{Beyond temporal localization}
	
	The proposed framework evaluates \emph{where} attributions concentrate in the cardiac cycle, a meaningful criterion as long as diagnostic guidelines directly specify the decisive segment. It reaches its natural boundary when the clinically relevant feature is not a temporal region but a morphological pattern such as the characteristic QRS shapes of the bundle branch blocks, the deviation and slope of the ST segment, or the compound duration of the PR interval. The bundle branch block histograms make this boundary concrete, as amplitude-sensitive methods collapse relevance onto the R-peak while SIGN-adjusted methods spread it along the QRS complex, a difference consistent with the diagnostic notion that QRS \emph{morphology} rather than amplitude distinguishes the two conditions, yet one that coverage cannot register because the mask spans the entire QRS region. Time-domain relevance maps assign scalar importance to individual samples and cannot represent such inter-sample relationships. Extending the framework towards concept-level attribution, for example via concept-wise relevance propagation \cite{Achtibat2023}, would allow explanations to be evaluated against morphological and interval-based features derived from the same guidelines, a natural next step.
	
	\subsection*{Limitations}
	
	Our analysis focused exclusively on gradient-based explanation methods, excluding perturbation-based, model-agnostic, and concept-based approaches to avoid the distortions associated with out-of-distribution perturbations \cite{Molnar2022,Schlegel2019}, although the framework itself can be extended to these paradigms. Within this scope, we assessed clinical plausibility rather than faithfulness, the model serving as a pattern proxy held constant across all methods (see above).

	The reference masks impose further boundaries, as they encode the dominant diagnostic cues but not every clinically relevant detail, which may underestimate methods that capture more complex or distributed patterns. They also encode only the temporal component of the criteria and are applied across all leads, so that relevance placed in the correct part of the cardiac cycle but in a clinically uninformative lead is still credited, but the lead focus can only be assessed via the lead-wise histograms. Relatedly, the spectral and coverage analyses evaluate only positive relevance, so that evidence encoded in negative relevance is not credited, which follows from the guideline designating only where positive evidence for a condition is expected. Since the masks occupy different fractions of the cardiac cycle, raw coverage is not comparable across pathologies, and all cross-pathology comparisons therefore rely on normalized coverage (NCov). The masks were derived from established guidelines and validated by cardiologists, though individual judgement may introduce minor boundary uncertainty, to which the mask sensitivity analysis suggests the rankings are robust.

	Finally, the empirical conclusions rest on a limited experimental base. The four pathologies cover principal categories of guideline criteria but not the full diagnostic spectrum of the ECG, as rhythm-based criteria and those relating successive beats are not represented, so that $NCov_{min}$ is a worst case over the conditions examined rather than over all conceivable ones. The models rest on a single CNN architecture trained on PTB-XL, and the observed amplitude-driven biases may vary for other architectures or datasets.
	
	\subsection*{Future directions}
	
	Future work should explore explanation techniques that embed clinical domain knowledge more directly, such as pattern- or prototype-based methods like ProtoPNet \cite{Chen2019}, and investigate guideline-aware constraints during training or explanation regularization. User-based evaluation is particularly important for clinical settings, where domain experts are the ultimate judges of explanation utility \cite{xaimedmanifesto,Ali2023}, yet it remains largely absent from ECG and time-series XAI, where evaluation still relies mainly on technical metrics or visual inspection rather than structured clinician judgment \cite{Salih2024,Manimaran2025,Gumpfer2026}. Closing this gap is a practical necessity rather than an academic preference, since a method that behaves well by technical criteria but is not judged useful by clinicians will not be adopted, and studies that let cardiologists rate, compare, or act on explanations, potentially guided by the reference regions established here, would test whether guideline-grounded quality translates into clinical utility. Extending the evaluation to additional model architectures, datasets, and pathologies, including rhythm-based criteria not covered here, would test how far the observed amplitude-driven biases generalize, and the framework itself is not restricted to gradient-based attribution, so it could likewise be applied to perturbation-based, model-agnostic, or concept-based explanation methods.

	\subsection*{Conclusion}
	
	Explainability in medical AI must be evaluated with the same rigor as predictive performance. Our results show that widely used explanation methods can introduce substantial and systematic biases when applied to ECG time series, and that these biases stay invisible under the local, sample-level inspection that currently dominates clinical AI research. Methods commonly applied in ECG analysis \cite{Manimaran2025,Goettling2024,Metsch2025,Strodthoff2019,Najjar2023,Guckert2021,Wagner2024,Strodthoff2021,AIME2024Paper}, among them Integrated Gradients, DeepSHAP, GradSHAP, and LRP-$\epsilon$, tie relevance to input magnitude through input multiplication or baseline subtraction and shift attribution toward high-amplitude waveform components, and nine of the 13 methods examined fell below chance on at least one pathology. Popularity is therefore a poor guide to reliability, and choosing an explanation method for ECG analysis calls for evidence across conditions of differing amplitude structure rather than convention or visual impression. Clinically plausible explanations therefore require methods that avoid amplitude-driven distortions, together with evaluation frameworks that can detect them before deployment. Domain-aware adaptations such as SIGN are one useful way to tailor gradient-based XAI to time-series data, accounting for signal-specific characteristics like amplitude semantics at minimal computational overhead. By making attribution behavior auditable against guideline-defined expectations, and by providing a pipeline extensible from method comparison to the analysis of individual models, the framework proposed here offers a pathway toward more dependable and clinically valid explanations for AI-assisted ECG analysis.
	
	\FloatBarrier
	\section*{Methods}
	
	\subsection*{Reference Patterns}
	\label{sec:reference_patterns}
	\FloatBarrier
	
	As a practice-oriented reference for evaluating XAI attribution patterns for ECG classification, we based our assessment on established medical guideline criteria \cite{AHA2008,AHA2009a,AHA2009b} for ECG interpretation. The analysis focused on four cardiac pathologies characterized by distinctive and diagnostically relevant ECG signatures. They were selected to cover the two categories into which guideline criteria for these signatures fall: criteria evaluated on low-amplitude intervals and segments close to the isoelectric line (the PR interval for AVB, the ST segment for ISCH) and criteria evaluated on high-amplitude QRS morphology (RBBB, LBBB). These categories account for a substantial share of the patterns used in routine ECG interpretation and, more importantly here, place opposite demands on an attribution method, so that a method biased toward signal amplitude can succeed on one category while failing on the other. These guideline-based reference patterns served as a gold standard to assess the extent to which model attributions reflected clinically meaningful features and aligned with expert reasoning. The temporal reference masks were derived through an iterative process involving clinical expert review, ensuring alignment with current diagnostic practice and providing a human-validated ground truth for the subsequent explanation evaluation. This approach directly responds to the identified need for domain-specific annotated benchmarks in XAI evaluation for time series.
	
	\subsubsection*{Atrioventricular Block} 
	
	An atrioventricular block (AVB) is a cardiac conduction disorder characterized by impaired signal transmission between the atria and ventricles \cite{AHA2008}. This impairment can cause delayed or completely blocked conduction, often resulting in a reduced heart rate. A key diagnostic criterion is prolongation of the interval between atrial depolarization (P) and ventricular depolarization (QRS) beyond 200\,ms in lead I or in other leads where the P-wave is clearly visible (Fig.~\ref{fig:patterns_criteria}). In Europe, this interval is commonly referred to as the \emph{PQ interval}, whereas in the United States it is called the \emph{PR interval} \cite{Houghton2019}. AVBs are classified from first degree (minimal delay) to third degree (complete block) according to the degree of conduction impairment \cite{AHA2008}.
	
	\subsubsection*{Myocardial Ischemia}  
	
	Myocardial ischemia (ISCH) is a condition in which the blood supply to the myocardium (heart muscle) is reduced, typically due to narrowed or blocked coronary arteries \cite{AHA2009b}. Ischemic regions produce characteristic ECG changes in anatomically related leads. The most important diagnostic criterion is the presence of ST-segment deviations, either elevation or depression, in one or more such leads. These changes are most often assessed in precordial leads (V1-V6), which correspond to anterior, lateral, and septal regions, thereby providing an approximate indication of the myocardial territories and coronary arteries that may be involved \cite{AHA2009b}.
	
	\subsubsection*{Right Bundle Branch Block} 
	
	A right bundle branch block (RBBB) is a conduction disorder characterized by delayed signal transmission in the right bundle branch of the cardiac conduction system, resulting in an abnormal intraventricular depolarization pattern \cite{AHA2009a}. RBBB can occur in a complete or incomplete form. The main ECG criteria of the complete form include a prolonged QRS duration ($>120$\,ms), an ``M''-shaped QRS complex in leads V1-V2, and a broad, slurred S-wave in leads V5-V6 (Fig.~\ref{fig:patterns_criteria}).
	
	\subsubsection*{Left Bundle Branch Block} 
	
	A left bundle branch block (LBBB) affects the left bundle branch of the cardiac conduction system \cite{AHA2009a}, which divides into a left anterior and a left posterior fascicle. Like RBBB, it may be complete or incomplete, and the two fascicles can also be blocked in isolation as left anterior or left posterior fascicular block, which is likewise identifiable on the ECG but by separate criteria and is not part of the LBBB label used here (see Data Source). Important leads for diagnosis are V1-V2 and V5-V6, too. The characteristic features of a complete LBBB are a prolonged QRS duration ($>120$\,ms), a deep S-wave in V1-V2, and a broad, notched R-wave in V5-V6 (Fig.~\ref{fig:patterns_criteria}).

	For both bundle branch blocks, complete and incomplete forms were summarized under a single label in this work (see Data Source). The criteria stated above are those of the complete forms, whereas the incomplete forms differ mainly in a shorter QRS duration (110-120\,ms) while presenting the same characteristic morphology in the same lead pairs \cite{AHA2009a}. The diagnostically relevant region is therefore the QRS complex in V1-V2 and V5-V6 in either case, so that one reference mask per pathology covers both forms.
	
	\medskip
	Based on the diagnostic criteria outlined above, we derived guideline-informed reference masks (Fig.~\ref{fig:patterns_criteria}) that delineate the regions of the cardiac cycle most likely to contain pathology-related patterns and distinguish them from regions of low diagnostic relevance. Boundaries were defined as fixed temporal segments within the standardized 500-timestep beat representation: the PR interval for AVB (P-wave onset to QRS onset), the ST segment for ISCH (QRS offset to T-wave onset), and the QRS complex for RBBB and LBBB. They were specified by two board-certified cardiologists and verified iteratively against median beat morphologies computed from the PTB-XL test set. Although the guidelines also specify the leads in which each pattern is assessed, the masks encode the \emph{temporal} component of the criteria only, and coverage is computed on relevance pooled across all 12 leads, so that the same mask applies to every lead and the masks for RBBB and LBBB coincide, both criteria residing on the QRS complex and differing in morphology and lead rather than in temporal position. The lead dimension remains visible in the lead-wise histograms (\SuppFigHist) but does not enter the coverage metric, which quantifies \emph{where in the cardiac cycle} relevance is placed.
	
	\subsection*{Data and Model}
	\subsubsection*{Data Source}
	
	We used the PTB-XL dataset \cite{Wagner2020a} (version 1.0.3) to train and evaluate four pathology-specific classification models. PTB-XL is an established public ECG dataset containing 21,799 12-lead ECG recordings of 10\,s length, sampled at 500\,Hz. The recordings are annotated with demographic information and up to 71 diagnostic statements in total, each accompanied by a confidence score (0-100\,\%). We excluded 426 recordings from patients younger than 18 years or with unknown age. Diagnostic labels were assigned according to the provided statements, applying a confidence threshold of $80\,\%$, which resulted in the removal of additional six records. For the two bundle branch blocks, complete and incomplete forms were summarized under the respective label (RBBB: CRBBB, IRBBB; LBBB: CLBBB, ILBBB), as both forms manifest with the same characteristic morphology in the same lead pairs and therefore share the same region of diagnostic interest (see Reference Patterns). Fascicular blocks (LAFB, LPFB) are annotated as separate statements in PTB-XL and were not counted as positives for LBBB. For each pathology, the negative class comprised all remaining eligible recordings, that is, those not carrying the respective positive label. From the remaining records, we used 30\,\% as a hold-out test set for each pathology. Of the remaining 70\,\%, 20\,\% were used for respective model validation and 80\,\% for training. All splits were stratified by sex and pathology label to preserve class distributions using a multi-label stratification approach \cite{Sechidis2011,Szymanski2017}. \SuppTabPerf\ provides an overview of the number of positive and negative cases for each pathology across the respective splits. PTB-XL is a publicly available, fully de-identified ECG dataset distributed through PhysioNet \cite{Goldberger2000}. As the present work relied solely on this existing open-access resource, no additional ethical approval was required for this study.
	
	\subsubsection*{Data Preprocessing}  
	
	We preprocessed the recordings using a 40\,Hz Butterworth low-pass filter, 50\,Hz and 0.05\,Hz second-order infinite impulse response (IIR) notch filters, and moving-average baseline correction. For model training, we extracted 4\,s subsamples (subsampling factor: 5) from each ECG using overlapping windows with uniform strides. This subsampling approach follows the process proposed alongside the model architecture described below \cite{GCB2020Paper}.
	
	\subsubsection*{Model Training}  
	
	We trained binary classification models using a CNN architecture optimized for short 12-lead ECG extracts \cite{GCB2020Paper}. Training was conducted on the splits listed in \SuppTabPerf\ using the Adam optimizer \cite{Kingma2015} with a learning rate of 0.001. To address class imbalance, we applied class-weight vectors \cite{Wang2016} during training. Model selection was based on maximizing the area under the receiver operating characteristic curve (AUROC) on the validation dataset while maintaining sensitivity and specificity above 0.7. The selected models were finally evaluated on the test set. All classifiers achieved high AUROC values (0.93-0.99) along with strong sensitivity (0.89-0.98) and specificity (0.82-0.98), see \SuppTabPerf\ for details. 
	
	\subsection*{Explanation Methods}\label{sec:explanationmethods}
	
	In this analysis, we focused on model-specific, local, post-hoc XAI methods that are commonly used in the field \cite{Manimaran2025,AIME2024Paper}, but often originate from the domain of computer vision \cite{Theissler2022,Wagner2024,Schlegel2019}. For detailed descriptions of the XAI methods used in this work, please refer to the \SuppMethods.
	
	\subsubsection*{Baseline Explanations}\label{sec:referencemethods}
	
	To establish reference points for evaluating explanation quality, we defined two baseline variants representing minimal semantic alignment with model behavior. First, a standard normal (Gaussian) distribution was used to generate an uninformative relevance map:
	\begin{equation}
		R \sim \mathcal{N}(0, 1)^{t \times 12},
	\end{equation}
	where $t$ denotes the temporal length of the input $x$. As a second baseline, the raw input signal itself was used as the relevance map:
	\begin{equation}
		R = x,
	\end{equation}
	representing an extreme case where relevance values directly correspond to input amplitudes.  
	
	Both baselines are expected to yield uninformative explanations, as the random map has no relation to model behavior while the input-based map reflects only signal magnitude rather than feature differentiation. If the raw input alone was sufficient to explain the model's predictions, the use of any additional explanation method would be redundant.
	
	\subsection*{Global Aggregation Pipeline}\label{sec:globalaggregationpipeline}
	
	The proposed evaluation framework treats local post-hoc explanations as building blocks for a global, pathology-level analysis. The pipeline consists of five steps that can be applied independently of the underlying explanation method or model architecture, and that are released as open-source code together with this study (see Data Availability).
	
	\textit{(1) Sample selection.} For each pathology, true-positive predictions are selected from the test set, ensuring that the analysis reflects explanations of correctly classified instances. \textit{(2) Local explanation generation.} For each selected sample, a local relevance map is generated using the explanation method under evaluation. \textit{(3) Beat extraction and alignment.} Two complete cardiac cycles are extracted from each recording on the basis of R-peak detection (see Beat Extraction below) and resampled to a fixed length of 500 timesteps to standardize temporal resolution across samples. \textit{(4) Normalization and filtering.} The positive part of each relevance map is normalized to the range $[0,1]$, and values below 0.05 are discarded to suppress background noise, with the rationale for evaluating positive relevance against the reference masks given under Pattern Analysis below. \textit{(5) Aggregation and comparison.} Normalized relevance maps are aggregated across all extracted beats to compute time-resolved histograms (see Spectral Analysis below), amplitude-relevance correlations (see Correlation Analysis below), and coverage of guideline-derived reference masks (see Coverage and Worst-case Reliability, and Reference Patterns below).
	
	The pipeline is method-agnostic and dataset-agnostic by design, requiring only a model, an explanation function, and a set of pathology-specific reference masks. It can therefore be extended to perturbation-based, model-agnostic, or concept-based explanation methods, and to other multivariate physiological time series for which guideline-derived reference masks can be defined.
	
	\subsection*{Beat Extraction}\label{sec:beatextractionandaggregation}
	
	For the analysis of explanation patterns, we focused on true-positive (TP) model predictions in the respective test sets. This choice ensures that explanations pertain to correctly classified instances, keeping the analysis focused on what the model has learned rather than on error modes, though this also means the framework cannot detect \emph{right-for-wrong-reasons} behavior on negative samples, which we acknowledge as a scope limitation. For each TP sample, local post-hoc explanations were generated using the methods described above. From each ECG recording and its corresponding relevance maps, two complete cardiac cycles were extracted. Selecting two beats per recording ensured a uniform sampling density across the dataset and provided adequate coverage under the assumption that each 4\,s window contained at least three QRS complexes. Recordings with fewer than three reliably detectable QRS complexes were excluded.
	
	Beat extraction was performed based on R-R intervals across all 12 leads. R-peak detection was implemented using the \textit{NeuroKit2} Python package \cite{Makowski2021neurokit}. Peaks were detected from the absolute signal of lead I, with the highest local maxima taken as QRS centers. Although this procedure does not yield clinically precise R-R intervals in all cases, it ensures consistent and reproducible beat-to-beat segmentation suitable for comparative analysis. For pathologies with abnormal QRS morphologies (RBBB, LBBB), R-peak detection from lead I may be less reliable, which can introduce minor misalignments in beat boundaries. The expected effect is a small reduction in histogram sharpness and a conservative bias toward lower coverage for these pathologies, which does not affect the qualitative ordering of methods. For each identified interval, the corresponding ECG signal and relevance values were extracted for all leads and explanation methods. Each beat cycle was subsequently resampled to a fixed length of 500 timesteps to standardize temporal resolution across samples.
	
	\subsection*{Pattern Analysis}\label{sec:patternanalysis}
	
	For all extracted beats, pathology-specific median beats and their interquartile ranges (0.25 and 0.75 quantiles) were computed on lead level \cite{Alcaraz2023,Wagner2024} (see top rows in \SuppFigHist). Because beats were extracted between consecutive QRS complexes, the leading and trailing portions of each interval were shifted to reconstruct complete cardiac cycles spanning from the P-wave to the T-wave. This approach is commonly applied in long-term ECG analysis to standardize beat alignment \cite{Jarchi2018}.
	
	For the spectral and coverage analyses, only positive relevance was considered, denoted $R_i^+ = \max(R_i, 0)$, and normalized per sample to the range $[0,1]$, with values below 0.05 discarded to suppress background noise. Restricting the analysis to positive relevance allowed us to isolate the features \textit{supporting} the predicted class and to compare them with the patterns specified in diagnostic guidelines. Since the analysis was further restricted to true-positive predictions (see Beat Extraction above), the relevant question was whether a model draws its support from the regions designated by the guideline. The correlation analysis below is an exception and uses $|R_i|$, since amplitude coupling concerns the strength of an attribution regardless of its direction. The resulting ECG beats and their corresponding normalized relevance maps formed the basis for the attribution pattern analysis and comparison with the guideline-informed reference masks (Fig.~\ref{fig:patterns_criteria}).
	
	\subsubsection*{Spectral Analysis}\label{sec:spectralanalysis}
	
	From the normalized and filtered relevance maps, time-resolved histograms were generated to characterize the distribution of relevance across the cardiac cycle (Fig.~\ref{fig:hist_summary} and \SuppFigHist). In these histograms, the $x$-axis represents the temporal position within the beat, the $y$-axis denotes the relevance magnitude, and the color scale encodes the relative frequency of occurrence. This representation enables the assessment of spatial distribution, intensity, and occurrence frequency of relevance values throughout each cardiac cycle. By visualizing the resulting relevance spectra for all methods, characteristic regions of consistently high or low relevance could be identified, highlighting systematic attribution patterns across beats and pathologies.
	
	\subsubsection*{Correlation Analysis}\label{sec:correlationanalysis}
	
	To examine potential dependencies between input magnitude and relevance assignment, we analyzed the relationship between absolute input values $|x_i|$ and corresponding absolute relevance values $|R_i|$ (Fig.~\ref{fig:hist_summary}). This relationship was quantified using Spearman's rank correlation coefficient (SCC) (see Table~\ref{tab:combined_analysis}), defined as
	\begin{equation}\label{eq:scc}
		\text{SCC}(x', R') = \frac{\sum_i (u(x')_i - \overline{u(x')})(v(R')_i - \overline{v(R')})}{\sqrt{\sum_i (u(x')_i - \overline{u(x')})^2} \sqrt{\sum_i (v(R')_i - \overline{v(R')})^2}},
	\end{equation}
	with
	\begin{align*}
		x' &:= \{\,|x_i| : x_i \in x\,\}, \\
		R' &:= \{\,|R_i| : R_i \in R\,\}, \\
		u(x') &:= \{\,r(x_i') : x_i' \in x'\,\}, \\
		v(R') &:= \{\,r(R_i') : R_i' \in R'\,\},
	\end{align*}
	where $u(x')$ and $v(R')$ represent the ascending ranks of the elements in $x'$ and $R'$, respectively. To ensure computational tractability across large test sets, the SCC was computed on a uniform random subsample of $N_s = 5 \times 10^5$ value pairs per method-pathology combination. After exclusion of non-finite pairs, the standard error of the correlation estimate remains below 0.005 and thus within the two-decimal rounding precision reported in Table~\ref{tab:combined_analysis}. This analysis enabled the quantification of amplitude bias, with higher correlation values indicating stronger dependence of relevance on input magnitude.
	
	\subsubsection*{Coverage and Worst-case Reliability}\label{sec:coverageconsistency}
	
	The \emph{coverage} metric quantifies the proportion of total relevance assigned to pathology-relevant regions as defined by the reference segmentation masks in Fig.~\ref{fig:patterns_criteria}. Coverage was computed at the population level, as the ratio of masked to total positive relevance accumulated over all analyzed beats and leads of a given method-pathology combination:
	\begin{equation}\label{eq:cov}
		Cov = \frac{\sum_{b} \sum_{l} \sum_t M_t \, R_{b,l,t}^+}{\sum_{b} \sum_{l} \sum_t R_{b,l,t}^+} \cdot 100,
	\end{equation}
	where $b$ indexes the extracted beats, $l$ the 12 leads, and $t$ the 500 timesteps of the standardized beat window, $M_t \in \{0,1\}$ denotes the binary mask indicating pathology-relevant segments, and $R_{b,l,t}^+$ the corresponding positive, normalized relevance (see Pattern Analysis above for why the analysis is restricted to the positive part). Coverage is therefore a single ratio of sums per method and pathology rather than an average of per-beat ratios, so that beats contribute in proportion to the total relevance they carry. Coverage values (in \%) are reported for each method and pathology in Table~\ref{tab:combined_analysis}.
	
	Raw coverage is not directly comparable across pathologies, because the masks occupy different fractions of the cardiac cycle by clinical convention, so a method assigning relevance at random already attains a non-zero, pathology-dependent coverage. To obtain a scale on which chance-level attribution maps to a common reference, we express coverage relative to the random baseline as a skill score. For pathology $i$, the \emph{normalized coverage} is
	\begin{equation}\label{eq:ncov}
		NCov_i = \frac{Cov_i - Cov^{\text{rand}}_i}{100 - Cov^{\text{rand}}_i} \cdot 100,
	\end{equation}
	where $Cov^{\text{rand}}_i$ is the coverage of the random baseline for pathology $i$ (Table~\ref{tab:combined_analysis}, Random row). By construction, $NCov_i = 0\,\%$ corresponds to chance-level attribution, $NCov_i = 100\,\%$ to perfect attribution of all relevance to the diagnostic region, and $NCov_i < 0\,\%$ to attribution that is \emph{worse} than chance, i.e., that systematically avoids the diagnostic region. Because the per-pathology baseline is subtracted, normalized coverage values are comparable across pathologies and are reported per pathology together with their mean $\overline{NCov}$. Coverage measures \emph{clinical plausibility}, rather than model faithfulness in a technical sense, as set out below.
	
	To summarize worst-case reliability, we further report the minimum normalized coverage across all examined pathologies:
	\begin{equation}\label{eq:cons}
		NCov_{min} = \min_{i \in \{1,\ldots,N\}} NCov_i,
	\end{equation}
	where $NCov_i$ denotes the normalized coverage (equation~\eqref{eq:ncov}) for pathology $i$. Where cross-pathology dispersion of a method is reported, it is the standard deviation of $NCov_i$ over the examined pathologies, computed on unrounded values and normalized by $N$ rather than $N-1$, since these pathologies constitute the complete set examined rather than a sample drawn from a larger population. Dispersion is reported on $NCov$ rather than on raw coverage, because the latter is not comparable across pathologies (see equation~\eqref{eq:ncov}). A high $NCov_{min}$ value indicates that a method maintains meaningful attribution even on its most challenging pathology, whereas a low or negative $NCov_{min}$ reveals that the method fails on at least one diagnostic condition regardless of its average performance. We adopt the minimum rather than the mean because an explanation method is chosen once and then applied to whichever condition a recording presents, so its practical reliability is bounded by its weakest condition, whereas averaging would allow strong agreement on one pattern category to mask systematic failure on the other.
	
	\subsection*{Clinical Plausibility vs. Faithfulness}\label{sec:scope}

	The framework measures \emph{clinical plausibility}, the agreement between an attribution and the guideline-defined diagnostic region, and deliberately does not measure \emph{faithfulness}. We apply no perturbation, deletion, or model-randomization test, and therefore make no claim about how tightly an individual method tracks the internal computation of the model. This is a scoping decision, as perturbation-based tests displace the input from the data manifold \cite{Molnar2022,Schlegel2019}, which is especially problematic for physiological signals whose diagnostic meaning resides in morphology, so that a drop in output confidence is not attributable to the removed evidence alone.

	Instead, the trained model serves as a \emph{pattern proxy}. The models are predictively valid on held-out data (AUROC 0.93-0.99, \SuppTabPerf), and the analysis is restricted to true-positive predictions, so on every sample examined the model reaches the correct conclusion. Prior work on the same architecture and pathologies \cite{AIME2024Paper} found the distribution of relevance across leads to agree broadly with the leads that guidelines prescribe for each condition. Most importantly, every explanation method is applied to the identical model for a given pathology, so any difference in coverage between two methods originates in the attribution rules rather than in what the model has learned. Where one method assigns substantial relevance to the ST segment while another assigns almost none, on the same model and the same recordings, the model cannot account for both outcomes at once.

	Two consequences follow for the interpretation of our results. A high coverage value indicates agreement between a method and the guideline, mediated by a model of established predictive validity, and is neither proof that the model itself reasons in guideline terms nor, for any single method, proof that its attribution is faithful. A coverage value read in isolation is correspondingly ambiguous between a model that ignores the region and a method that fails to expose it. What the shared model licenses is instead the \emph{relative} statement, in that methods which disagree on a fixed model and fixed recordings cannot all be tracking it equally well, so that the ranking between them is informative even where the absolute level is not. Establishing the method-model link itself would require faithfulness tests complementary to, and outside the scope of, the evaluation proposed here.

	\subsection*{Sensitivity and Uncertainty Analysis}\label{sec:robustness_methods}
	
	\paragraph{Bootstrap confidence intervals.} To quantify the sampling uncertainty of the coverage estimates, we computed 95\,\% bootstrap confidence intervals by resampling the set of true-positive test recordings with replacement (1000 draws per method-pathology combination), as reported in \SuppTabBoot. The population-level coverage per draw was computed as the ratio of total masked relevance to total relevance across the resampled recordings, matching the metric used in the main analysis. The random baseline coverage used for NCov normalisation was held fixed at its observed mean (not resampled), since it characterizes the mask geometry rather than the method under test. For $NCov_{min}$, the minimum NCov across pathologies was computed within each bootstrap draw before taking the 2.5th and 97.5th percentiles.
	
	\paragraph{Mask boundary sensitivity.} To assess whether the results depend on the exact mask boundaries, we repeated the coverage analysis under symmetric expansions and contractions of the mask edges by $k \in \{-20, -10, 0, +10, +20\}$~timesteps (where one timestep corresponds to 1/500 of the standardized beat window), with $NCov_{min}$ per method reported in \SuppTabMask. At each offset, the mask was derived from the baseline definition by binary dilation ($k > 0$) or erosion ($k < 0$) of each contiguous masked region, and the random baseline was recomputed with the same shifted mask to maintain a consistent NCov normalisation. All other pipeline steps were identical to the main analysis.

	\section*{Data Availability}
	
	All code required to reproduce the experiments and analyses presented in this study, including the proposed global, guideline-grounded evaluation pipeline, is publicly available at {\small\textbf{\url{https://github.com/nilsgumpfer/globalxaiecg}}}. The ECG records and pathology labels used in this work were obtained from the PTB-XL dataset \cite{Wagner2020a}, which is openly accessible via PhysioNet \cite{Goldberger2000} at {\small\textbf{\url{https://physionet.org/content/ptb-xl/1.0.3}}}. Both the code repository and the dataset are freely available for research purposes under their respective licenses.
	
	\bibliography{literature}
	
	\section*{Acknowledgements}
	
	This work was supported by the German Federal Ministry of Research, Technology and Space (BMFTR) through ExperTeam4KI (grant no.~16IS24063). We gratefully acknowledge support from the hessian.AI Service Center (funded by the BMFTR, grant no.~16IS22091) and the hessian.AI Innovation Lab (funded by the Hessian Ministry for Digital Strategy and Innovation, grant no.~S-DIW04/0013/003).
	
	\section*{Author contributions statement}
	
	All authors contributed to the conceptualization of the study. N.G., M.G., and J.H. developed the methodology. N.G. curated the data, implemented the software, performed the experiments and analysis, and visualized the results. N.G. prepared the original draft of the manuscript, and all authors contributed to writing, reviewing, and editing. Medical guidance was provided by B.A. and S.S., while the overall study was supervised by J.H.
	
	\section*{Additional information}
	
	\textbf{Competing interests:} The authors declare no competing financial interests. We disclose that the SIGN-based methods evaluated in this study were proposed in own prior work~\cite{Gumpfer2023}. All methods were compared under an identical, openly released, pre-specified pipeline, so that no method, including the authors' own, received preferential treatment.
	
	\FloatBarrier

\end{document}